\documentclass[epj]{svjour}
\usepackage{graphics}

\def\be{\begin{eqnarray}}
\def\ee{\end{eqnarray}}

\begin{document}

\title{Light flavor asymmetry of nucleon sea}
\author{Huiying~Song\inst{1} \and Xinyu~Zhang\inst{1} \and Bo-Qiang~Ma\inst{1}$^{,}$\inst{2}$^{,}$\thanks{Corresponding author. Email address: \texttt{mabq@pku.edu.cn}}
}
%
%
\institute{School of Physics and State Key Laboratory of Nuclear Physics and
Technology, Peking University, Beijing 100871, China \and Center for High Energy
Physics, Peking University, Beijing 100871, China}
%
\date{Received: date / Revised version: date}
%
\abstract{ The light flavor antiquark distributions of the nucleon
sea are calculated in the effective chiral quark model and compared
with experimental results. The contributions of the flavor-symmetric
sea-quark distributions and the nuclear EMC effect are taken into
account to obtain the ratio of Drell-Yan cross sections
$\sigma^{\mathrm{pD}}/2\sigma^{\mathrm{pp}}$, which can match well
with the results measured in the FermiLab E866/NuSea experiment. The
calculated results also match the measured $\bar{d}(x)-\bar{u}(x)$
from different experiments, but unmatch the behavior of
$\bar{d}(x)/\bar{u}(x)$ derived indirectly from the measurable quantity
$\sigma^{\mathrm{pD}}/2\sigma^{\mathrm{pp}}$ by the FermiLab
E866/NuSea Collaboration at large $x$. We suggest to measure again
$\bar{d}(x)/\bar{u}(x)$ at large $x$ from precision experiments with
careful experimental data treatment. We also propose an alternative
procedure for experimental data treatment.
%
}
\titlerunning{Light flavor asymmetry of nucleon sea}
\authorrunning{Huiying~Song {\it et. al.}}
\maketitle

\section{INTRODUCTION}

The nucleon sea is an active issue in hadron physics because of its
importance in understanding both the nucleon structure and
properties of strong interaction. In the early days, it was usually
assumed that the sea of the proton was flavor symmetric between
$u\bar{u}$ and $d\bar{d}$ quark-antiquark pairs, i.e.,
$\bar{u}^{\mathrm{p}}(x)=\bar{d}^{\mathrm{p}}(x)$. However, this
assumption was found to be unjustified by the observation of the
Gottfried sum rule (GSR)~\cite{gottfried1967} violation from a
number of
experiments~\cite{pa1991,nmc1994,NA51,E8661998,pea98,E866,HERMES}.
There have been many studies related to theoretical explanations of
these observations and the flavor asymmetry of the nucleon
sea~\cite{Rev}.

The Gottfried sum~\cite{gottfried1967} is defined as
$S_{\mathrm{G}}=\int_{0}^{1}[F_{2}^{\mathrm{p}}(x)-F_{2}^{\mathrm{n}}(x)]{\mathrm{d}x/x}$,
which, when expressed in terms of quark momentum distributions,
takes the form
\begin{eqnarray}
S_{\mathrm{G}}&=&\int_{0}^{1}[F_{2}^{\mathrm{p}}(x)-F_{2}^{\mathrm{n}}(x)]\frac{\mathrm{d}x}{x}\nonumber\\
&=&\int_{0}^{1}\sum _{i}
e_{i}^{2}[q_{i}^{\mathrm{p}}(x)+\bar{q}_{i}^{\mathrm{p}}(x)-q_{i}^{\mathrm{n}}(x)-\bar{q}_{i}^{\mathrm{n}}(x)]
\mathrm{d}x,
\end{eqnarray}
where $x$ is the Bjorken variable, $F_{2}^{\mathrm{p}}(x)$ and
$F_{2}^{\mathrm{n}}(x)$ are the proton and the neutron structure
functions, $e_{i}$ is the charge of the quark of flavor $i$, and
$q_{i}^{\mathrm{N}}$~($\bar{q}_{i}^{\mathrm{N}}$, N=n,p) is the
quark (antiquark) distribution of the nucleon. Decomposing the quark
distribution into valance ($V$) and sea ($Q$ for quark and $A$ for
antiquark) components, we get \be
\int_{0}^{1}q_{i}^{\mathrm{N}}(x)\mathrm{d}x=\int_{0}^{1}[V_{i}^{\mathrm{N}}(x)+Q_{i}^{\mathrm{N}}(x)]\mathrm{d}x,
\ee \be
\int_{0}^{1}\bar{q}_{i}^{\mathrm{N}}(x)dx=\int_{0}^{1}A_{i}^{\mathrm{N}}(x).
\ee From the flavor number conservation and the isospin symmetry between
the proton and the neutron, i.e.,
$u^{\mathrm{p}}(x)=d^{\mathrm{n}}(x)$,
$\bar{u}^{\mathrm{p}}(x)=\bar{d}^{\mathrm{n}}(x)$, etc., we get \be
S_{\mathrm{G}}=\frac{1}{3}-\frac{2}{3}\int_{0}^{1}[\bar{d}^{\mathrm{p}}(x)-\bar{u}^{\mathrm{p}}(x)]\mathrm{d}x.
\ee If the nucleon sea is flavor symmetric, one arrives at the GSR
\be S_{\mathrm{G}}=\frac{1}{3}. \ee

The violation of the GSR was first observed by the New Muon
Collaboration (NMC)~\cite{pa1991} at CERN in 1991, and the reported
Gottfried sum is
$\int_{0}^{1}[F_{2}^{\mathrm{p}}(x)-F_{2}^{\mathrm{n}}(x)]{\mathrm{d}x/x}=0.240\pm
0.016$. In 1994 the NMC reanalyzed their data~\cite{nmc1994}, and
got
$\int_{0}^{1}[F_{2}^{\mathrm{p}}(x)-F_{2}^{\mathrm{n}}(x)]{\mathrm{d}x/x}=0.235\pm
0.026$. Though the isospin symmetry breaking between the proton and the
neutron at the parton level could also contribute~\cite{m92}, at
least partially, to the GSR violation, the results are usually
interpreted as an indication of light flavor asymmetry of the
nucleon sea~\cite{Preparata:1990fd}, and the excess of $d\bar{d}$
pairs over $u\bar{u}$ pairs in the nucleon sea was measured
as~\cite{nmc1994}
 \be
\int_{0}^{1}[\bar{d}^{\mathrm{p}}(x)-\bar{u}^{\mathrm{p}}(x)]\mathrm{d}x=0.148\pm
0.039. \ee

The Drell-Yan process~\cite{Drell1971} can be used to measure the
flavor distribution of the nucleon sea. The cross section of the
Drell-Yan process at leading order is
 \begin{eqnarray}
 \sigma^{\mathrm{AB}} \propto
 \sum_{i}e_{i}^{2}[q_{i}^{\mathrm{A}}(x_{1},Q^{2})\bar{q}_{i}^{\mathrm{B}}(x_{2},Q^{2})\nonumber\\ +\bar{q}_{i}^{\mathrm{A}}(x_{1},Q^{2})q_{i}^{\mathrm{B}}(x_{2},Q^{2})],
\label{sigma}
\end{eqnarray}
where the sum is over all quark flavors, $e_{i}$ is the charge of
the quark of flavor $i$,
$q_{i}^{\mathrm{A}}~(\bar{q}_{i}^{\mathrm{A}})$ is the quark
(antiquark) distribution of the beam,
$q_{i}^{\mathrm{B}}~(\bar{q}_{i}^{\mathrm{B}})$ is the quark
(antiquark) distribution of the target, and  $x_{1}$ and $x_{2}$ are
the Bjorken variables $x$ of the partons from the beam and the
target respectively. Two kinematic quantities commonly used to
describe Drell-Yan process are Feynman-$x$,
$x_{\mathrm{F}}=x_{1}-x_{2}$, and the dilepton mass $M$,
$M^2=Q^2\approx x_{1}x_{2}s$, where $Q^2$ and $\sqrt{s}$ are the
square of the invariant momentum transferred and the center-of-mass
energy of the initial nucleons respectively.

The FermiLab E772 Collaboration~\cite{E772} reported an upper limit
on the $\bar{d}$-$\bar{u}$ asymmetry in the range $0.04\leq x \leq
0.27$. Later, the CERN NA51 experiment~\cite{NA51} measured the
ratio of cross sections for muon pair production through the
Drell-Yan process in $\mathrm{p}\mathrm{p}$ and
$\mathrm{p}\mathrm{D}$ reactions at $y \sim  0$, with $450$~GeV/c
incident protons. The Drell-Yan asymmetry is measured as
$A_{\rm{DY}}= (\sigma^{\rm{pp}}-\sigma^{\rm{pn}}) /
(\sigma^{\rm{pp}}+\sigma^{\rm{pn}}) =-0.09\pm0.02\pm0.025$. The
ratio of $\bar{u}$ over $\bar{d}$ of the nucleon sea derived from
this measurement is \be
\frac{\bar{u}^{\mathrm{p}}}{\bar{d}^{\mathrm{p}}}|_{<x>=0.18}=0.51\pm0.04\pm0.05.
\ee However, the acceptance of NA51 spectrometer is peaked near
$x_{\mathrm{F}}=0$ and $x=0.18$, and consequently we are almost
impossible to determine the $x$-dependence of $\bar{d}^{\mathrm{p}}
/ \bar{u}^{\mathrm{p}}$ based on the NA51 experiment.

The FermiLab E866/NuSea Collaboration~\cite{E866} measured the ratio
of cross sections of Drell-Yan muon pairs from $800$~GeV/c proton
beam scattered on liquid hydrogen and deuterium targets and
extracted $\bar{d}(x) / \bar{u}(x)$ and $\bar{d}(x)-\bar{u}(x)$ in
the proton sea over a wide range of $x$, and got
$\int_{0}^{1}[\bar{d}(x)-\bar{u}(x)]\mathrm{d}x=0.118\pm0.012$.
However, the shape of $\bar{d}^{\mathrm{p}}(x) /
\bar{u}^{\mathrm{p}}(x)$ is beyond expectation and the ratio can
even be less than 1 when $x$ is large, which has received widespread
attention because so large flavor asymmetry of $\bar{d}$ and
$\bar{u}$ was unexpected and it is even more difficult to explain
the result that $\bar{d}(x)<\bar{u}(x)$ when $x$ is large. The
HERMES Collaboration~\cite{HERMES} measured charged hadrons from
semi-inclusive deep-inelastic scattering, and reported
$\bar{d}(x)-\bar{u}(x)$ over the range $0.02\leq x \leq 0.3$ and
$1~\mathrm{GeV}^{2}/\mathrm{c}^{2} < Q^{2} <
10~\mathrm{GeV}^{2}/\mathrm{c}^{2}$. Their result of
$\bar{d}(x)-\bar{u}(x)$ is in agreement with that reported by the
E866/NuSea experiment.

In 1977 Field and Feynman~\cite{Field} pointed out that
$\bar{d}=\bar{u}$ would not strictly hold even in the perturbative
Quantum Chromodynamics (QCD), due to the fact that an extra valance
up quark in the proton can lead to a suppression of $g\rightarrow
u\bar{u}$ relative to $g\rightarrow d\bar{d}$ via Pauli blocking.
Nevertheless later calculations~\cite{Ross,Steffens} indicated that
the effects of Pauli blocking are very small and such large
asymmetry must have a nonperturbative origin. The role of mesons in
DIS was first investigated by Sullivan~\cite{Sullivan}. He suggested
that some fraction of the sea antiquark distribution of the nucleon
may be associated with the pion cloud around the nucleon core. This
was the original idea of the meson cloud model. Many authors used
the pion cloud mechanism~\cite{Thom83,Kuma91,emh,as,js,wyp} or
baryon-meson fluctuation picture~\cite{Bo-Qiang Ma} of the nucleon
to explain the light flavor asymmetry of the nucleon sea. Besides,
the effective chiral quark model~\cite{Weinberg,mg84} is also a
method to explain the nucleon sea flavor
asymmetry~\cite{Eich92,wakamatsu} and we discuss the details in
Sec.~\ref{section2}. There are many other mechanisms to explain
$\bar{d}\neq\bar{u}$, for example, chiral soliton
model~\cite{Pobylitsa}, instanton model~\cite{Dorokhov}, statistical
model~\cite{statismodel} and so on. But until now, no model can
explain the $\bar{d}(x)/\bar{u}(x)<1$ behavior in the large $x$
region.

In Sec.~\ref{section2}, we use the effective chiral quark
model~\cite{Weinberg,mg84}, with the constituent quark
model~\cite{hz81} and the light-cone quark-spectator-diquark
model~\cite{parton} results respectively as the bare constituent
quark distributions inputs, to calculate the quark distributions of
the proton. From the isospin symmetry between the proton and the
neutron, we obtain the quark distributions of the neutron. In
Sec.~\ref{section3}, we take the symmetric quark and antiquark sea
contribution and the $Q^{2}$-evolution of quark distribution into
consideration, to obtain
$\sigma^{\mathrm{pD}}/2\sigma^{\mathrm{pp}}$,
$\bar{d}(x)-\bar{u}(x)$ and $\bar{d}(x)/\bar{u}(x)$. In
Sec.~\ref{section4}, we discuss the possible nuclear EMC effect in
the extraction of the ratio $\bar{d}(x)/\bar{u}(x)$. By taking into
account such effect, the behavior of
$\sigma^{\mathrm{pD}}/2\sigma^{\mathrm{pp}}$, which are really
measured quantities rather than $\bar{d}(x)/\bar{u}(x)$, can match
with the experimental results better when $x$ is large.
Sec.~\ref{section5} is devoted to some conclusions and summary.

\section{THE SEA CONTENT IN THE EFFECTIVE CHIRAL QUARK MODEL}\label{section2}

The effective chiral quark model, established by
Weinberg~\cite{Weinberg}, and developed by Manohar and
Georgi~\cite{mg84}, has been widely recognized by the hadron physics
society as an effective theory of QCD at the low energy scale. The
effective chiral quark model has an apt description of its important
degrees of freedom in terms of quarks, gluons and Goldstone~(GS)
bosons at momentum scales relating to hadron structure. There has
been a prevailing impression that the effective chiral quark model
is successful in explaining the violation of GSR from a microscopic
viewpoint~\cite{Eich92,wakamatsu}. Also, this model plays an
important role in explaining the proton spin crisis~\cite{a89} in
Refs.~\cite{cl95,songxiaotong}. A study by Ding and Xu with one of
us~\cite{dxm04} also shows that the strange-antistrange asymmetry
within the effective chiral quark model could explain the NuTeV
anomaly. We adopt the effective chiral quark model to calculate the
quark and antiquark distributions of nucleons in this paper. The
principles and basic formulas are almost the same as those used
previously, but the options and inputs are carefully considered and
some of them are differently chosen to make the results closer to
the data.

The chiral symmetry at the high energy scale and its breaking at the low
energy scale are the basic properties of QCD. Because the effect of
the internal gluons is small in the effective chiral quark model at the
low energy scale, the gluonic degrees of freedom are negligible when
comparing to GS bosons and quarks. In this picture, the
valence quarks contained in the nucleon fluctuate into quarks plus
GS bosons, which spontaneously break chiral symmetry, and any low
energy hadron properties should include this symmetry violation. The
effective interaction Lagrangian is
\begin{equation}
L=\bar{\psi}(iD_{\mu}+V_{\mu})\gamma^{\mu}\psi+ig_{A}\bar{\psi}A_{\mu}\gamma^{\mu}\gamma_{5}\psi+\cdots,
\end{equation}
where
\begin{equation}
\psi=\left(%
\begin{array}{c}
  u \\
  d \\
  s \\
\end{array}%
\right)
\end{equation}
is the quark field and $D_{\mu}=\partial_{\mu}+igG_{\mu}$ is the
gauge-covariant derivative of QCD, with $G_{\mu}$ standing for the
gluon field, $g$ standing for the strong coupling constant and
$g_{A}$ standing for the axial-vector coupling constant determined from the
axial charge of the nucleon. $V_{\mu}$ and $A_{\mu}$ are the vector
and the axial-vector currents which are defined by
\begin{equation}
\left(%
\begin{array}{c}
  V_{\mu} \\
  A_{\mu} \\
\end{array}%
\right)=\frac{1}{2}(\xi^{+}\partial_{\mu}\xi\pm\xi\partial_{\mu}\xi^{+}),
\end{equation}
where $\xi=\mathrm{exp}(i\Pi/f)$, and $\Pi$ has the form
\begin{equation}
\Pi\equiv\frac{1}{\sqrt{2}}\left(
\begin{array}{ccc}
  \frac{\pi^{0}}{\sqrt{2}}+\frac{\eta}{\sqrt{6}} & \pi^{+} & K^{+} \\
  \pi^{-} & -\frac{\pi^{0}}{\sqrt{2}}+\frac{\eta}{\sqrt{6}} & K^{0} \\
  K^{-} & \overline{K^{0}} & \frac{-2\eta}{\sqrt{6}} \\
\end{array}
\right).
\end{equation}
With the expansions for $V_{\mu}$ and $A_{\mu}$ in powers of $\Pi/f$,
it gives $V_{\mu}=0+O(\Pi/f)^{2}$ and
$A_{\mu}=i\partial_{\mu}\Pi/f+O(\Pi/f)^{2}$, where the pseudoscalar
decay constant is $f\simeq93$~MeV. Thus, the effective interaction
Lagrangian between GS bosons and quarks in the leading order
becomes~\cite{Eich92}
\begin{equation}
L_{\Pi
q}=-\frac{g_{A}}{f}\bar{\psi}\partial_{\mu}\Pi\gamma^{\mu}\gamma_{5}\psi.
\end{equation}
We should point out that we use the perturbative expansion in the
energy rather than in the effective couple constant, which can be
large. Although this model contains an infinite number of terms, at
a given order in the energy expansion, the low-energy theory is
specified by a finite number of couplings. Therefore if the energy
scale that we consider is low, the perturbative expansion is
applicable regardless of the value of the coupling constant. The
framework that we use in this paper is based on the time-ordered
perturbative theory in the infinite momentum frame~(IMF). Because
all particles are on-mass-shell in this frame and the factorization
of the subprocess is automatic, we neglect all possible
off-mass-shell corrections. In this framework, we can express the
quark distributions inside a nucleon as a convolution of a
constituent quark distribution in a nucleon and the structure
functions of a constituent quark. The light-front Fock
decompositions of constituent quark wave functions have the
following forms
\begin{equation}
|U\rangle=\sqrt{Z}|u_{0}\rangle+a_{\pi}|d\pi^{+}\rangle+\frac{a_{\pi}}{%
\sqrt{2}}|u\pi^{0}\rangle+a_{K}|sK^{+}\rangle+\frac{a_{\eta}}{\sqrt{6}}%
|u\eta\rangle,\label{u}
\end{equation}
\begin{equation}
|D\rangle=\sqrt{Z}|d_{0}\rangle+a_{\pi}|u\pi^{-}\rangle+\frac{a_{\pi}}{%
\sqrt{2}}|d\pi^{0}\rangle+a_{K}|sK^{0}\rangle+\frac{a_{\eta}}{\sqrt{6}}%
|d\eta\rangle.\label{d}
\end{equation}
Here, $Z$ is the renormalization constant for the bare constituent
quarks which are massive and denoted by $|u_{0}\rangle$ and
$|d_{0}\rangle$, and $|a_{\alpha}|^{2}$ are the probabilities to
find GS bosons in the dressed constituent quark states ($|U\rangle$
for an $up$ quark and $|D\rangle$ for a $down$ quark), where
$\alpha=\pi, K, \eta$. In the effective chiral quark model, the
fluctuation of a bare constituent quark into a GS boson and a recoil
bare constituent quark can be given as~\cite{sw98}
\begin{equation}
q_{j}(x)=\int^{1}_{x}\frac{\textmd{d}y}{y}P_{j\alpha/i}(y)q_{i}\left(\frac{x}{y}\right).\label{q}
\end{equation}
In Eq.~(\ref{q}), $P_{j\alpha/i}(y)$ is the splitting function for
the probability to find a constituent quark $j$ carrying the
light-cone momentum fraction $y$ together with a spectator GS
boson~$\alpha$, and it has the following form
\begin{eqnarray}
P_{j\alpha/i}(y)&=&\frac{1}{8\pi^{2}}\left(\frac{g_{A}\overline{m}}{f}\right)^{2}\nonumber\\
&\times&\int
\textmd{d}k^{2}_{T}\frac{(m_{j}-m_{i}y)^{2}+k^{2}_{T}}{y^{2}(1-y)[m_{i}^{2}-M^{2}_{j\alpha}]^{2}},
\label{splitting}
\end{eqnarray}
where $m_{i}, m_{j}, m_{\alpha}$ are the masses of the $i$-, $
j$-constituent quarks and the pseudoscalar meson $\alpha$,
respectively, $\overline{m}=(m_{i}+m_{j})/2$ is the average mass of
constituent quarks, and $M_{j\alpha}^{2}$ is the square of the invariant mass of the final states,
\begin{equation}
M^{2}_{j\alpha}=\frac{m^{2}_{j}+k^{2}_{T}}{y}+\frac{m^{2}_{\alpha}+k^{2}_{T}}{1-y}.
\end{equation}

In this paper, we adopt the definition of the moment of the
splitting function
\begin{equation}
\langle x^{n-1}
P_{j\alpha/i}\rangle=\int^{1}_{0}x^{n-1}P_{j\alpha/i}(x)\textmd{d}x
\end{equation}
with the first moment $\langle P_{j\alpha/i}\rangle=\langle
P_{\alpha j/i}\rangle\equiv \langle
P_{\alpha}\rangle=|a_{\alpha}|^{2}$~\cite{sw98}. In terms of the
above notation, the renormalization constant $Z$ is given by
\begin{equation}
Z=1-\frac{3}{2}\langle P_{\pi}\rangle-\langle P_{K}\rangle-\frac{1}{6}\langle P_{\eta}\rangle.
\end{equation}

Now we need to specify the momentum cutoff function at the quark-GS
boson vertex. It is conventional to use an exponential cutoff
in IMF calculations,
\begin{equation}
g_{A}\rightarrow
g_{A}^{\prime}\textmd{exp}\bigg{[}\frac{m^{2}_{i}-M^{2}_{j\alpha}}{4\Lambda^{2}}\bigg{]},
\end{equation}
with $g_{A}^{\prime}=1$ following the large $N_{c}$
argument~\cite{w90}. However, $g_{A}^{\prime}=0.75$ was adopted in
the original work~\cite{mg84}. Such a form factor has the correct
$t$ and $u$ channel symmetry, and $\Lambda$ is the cutoff parameter,
which is determined by the experimental data of the Gottfried sum
and the constituent quark mass inputs for the pion. This function
satisfies the symmetry $P_{j \alpha /i} (y) = P_{\alpha j/i} (1-y)$.

When probing the internal structure of GS bosons, we can write
the process in the following form~\cite{sw98}
\begin{equation}
q_{k}(x)=\int\frac{\textmd{d}y_{1}}{y_{1}}\frac{\textmd{d}
y_{2}}{y_{2}}V_{k/\alpha}\left(\frac{x}{y_{1}}\right)P_{\alpha
j/i}\left(\frac{y_{1}}{y_{2}}\right)q_{i}\left(y_{2}\right),
\end{equation}
where $V_{k/\alpha}(x)$ is the quark $k$ distribution function in
$\alpha$ and satisfies the normalization
$\int_{0}^{1}V_{k/\alpha}(x)dx=1$. Because the mass of $\eta$ is so
high and the coefficient is so small that the fluctuation of it is
suppressed, the contribution of $\eta$ is not considered in our
calculation. From Eqs.~(\ref{u}) and (\ref{d}), we obtain the quark distribution functions of nucleon by using the
splitting function Eq.~(\ref{splitting}) and the constituent quark
distributions $u_{0}$ and $d_{0}$,
\begin{eqnarray}
u(x)&=&Zu_{0}(x)+P_{u\pi^{-}/d}\otimes d_{0}+V_{u/\pi^{+}}\otimes
P_{\pi^{+}d/u}\otimes u_{0}\nonumber\\
&+&\frac{1}{2}P_{u\pi^{0}/u}\otimes
u_{0}+V_{u/K^{+}}\otimes P_{K^{+}s/u}\otimes u_{0}\nonumber\\
&+&
\frac{1}{2}V_{u/\pi^{0}}\otimes (P_{\pi^{0}u/u}\otimes u_{0}+P_{\pi^{0}d/d}\otimes d_{0}),\nonumber\\
d(x)&=&Zd_{0}(x)+P_{d\pi^{+}/u}\otimes u_{0}+V_{d/\pi^{-}}\otimes
P_{\pi^{-}u/d }\otimes d_{0}\nonumber\\
&+& \frac{1}{2}P_{d\pi^{0}/d}\otimes
d_{0}+V_{d/K^{0}}\otimes P_{K^{0}s/d}\otimes d_{0}\nonumber\\
&+&\frac{1}{2}V_{d/\pi^{0}}\otimes (P_{\pi^{0}u/u }\otimes
u_{0}+P_{\pi^{0}d/d}\otimes d_{0}).~\label{uddis}
\end{eqnarray}
Here, we define the notations for the convolution integral as
\begin{equation}
P_{j \alpha / i}\otimes q_i =\int_{x}^{1}\frac{\textmd{d}y}{y}P_{j
\alpha / i}\left(y\right)q_i\left(\frac{x}{y}\right),
\end{equation}
and
\begin{eqnarray}
V_{k/ \alpha}&\otimes &P_{\alpha j/i}\otimes
q_i=\nonumber\\
\int_{x}^{1}&\frac{\textmd{d}y_{1}}{y_{1}}&\int_{y_{1}}^{1}\frac{\textmd{d}y_{2}}{y_{2}}V_{k/
\alpha}\left( \frac{x}{y_{1}}\right)P_{\alpha
j/i}\left(\frac{y_{1}}{y_{2}}\right)q_{i}\left(y_{2}\right).
\end{eqnarray}
In the same way, we can derive the light-flavor antiquark
distributions,
\begin{eqnarray}
\bar{u}(x)&=&V_{\bar{u}/\pi^{-}}\otimes P_{\pi^{-}u/d}\otimes
d_{0}\nonumber\\&+&\frac{1}{2}V_{\bar{u}/\pi^{0}}\otimes (P_{\pi^{0}u/u}\otimes
u_{0}+P_{\pi^{0}d/d}\otimes d_{0}),\nonumber\\
\bar{d}(x)&=&V_{\bar{d}/\pi^{+}}\otimes P_{\pi^{+}d/u}\otimes
u_{0}\nonumber\\&+&\frac{1}{2}V_{\bar{d}/\pi^{0}}\otimes (P_{\pi^{0}u/u}\otimes
u_{0}+P_{\pi^{0}d/d}\otimes d_{0}),\nonumber\\
\label{udsdis}
\end{eqnarray}
where \begin{eqnarray}
&&~~~V_{u/\pi^{+}}=V_{\bar{d}/\pi^{+}}=V_{d/\pi^{-}}=V_{\bar{u}/\pi^{-}}\nonumber\\&&=2V_{u/\pi^{0}}
=2V_{\bar{u}/\pi^{0}}=2V_{d/\pi^{0}}=2V_{\bar{d}/\pi^{0}}
\nonumber\\&&=\frac{1}{2}V_{\pi}\left(x\right),
\end{eqnarray} and
$$V_{u/K^{+}}=V_{d/K^{0}}.$$
From above equations, we can reexamine the valence quark
distributions $u_{v}(x)=u(x)-\bar{u}(x)$ and
$d_{v}(x)=d(x)-\bar{d}(x)$, which satisfy the correct normalization
with the renormalization constant $Z$. We should point out that in
the chiral quark model, antiquarks are produced in the process of
the splitting of Goldstone bosons unless higher order effects are
considered. Since Goldstone bosons are spin-0 particles without
polarization, antiquarks are unpolarized, and this feature is
compatible with the available experimental data~\cite{Hermes}.

In this paper, we choose $m_{u}=m_{d}=330$~MeV, $m_{s}=480$~MeV,
$m_{\pi^{\pm}}=m_{\pi^{0}}=140$~MeV and
$m_{K^{+}}=m_{K^{0}}=495$~MeV. Employing the quark distributions of
the effective chiral quark model, we get the Gottfried sum
determined by the difference between the proton and the neutron
structure functions,
\begin{eqnarray}
S_{G}&=&\int^{1}_{0}\frac{\mathrm{d}x}{x}[F^{p}_{2}(x)-F^{n}_{2}(x)]\nonumber\\
&=&\frac{1}{3}\int_{0}^{1}\mathrm{d}x[u(x)+\bar{u}(x)-d(x)-\bar{d}(x)]\nonumber\\&
=&\frac{1}{3}(Z-\frac{1}{2}\left<P_{\pi}\right>+\left<P_{K}\right>+\frac{1}{6}\left<P_{\eta}\right>)\nonumber\\
&=&\frac{1}{3}(1-2\left<P_{\pi}\right>).\label{gottfried}
\end{eqnarray}
From the above equation and the experimental value of Gottfried
sum~\cite{nmc1994}, we can find that the appropriate value for
$\Lambda_{\pi}$ is $1500$~MeV.  At the same time, $\langle
P_{\pi}\rangle=0.149$, $\langle P_{K}\rangle=0.085$, $\langle
P_{\eta}\rangle=0.063$ and $Z=0.682$. But for $K$ and $\eta$ mesons,
the terms $\langle P_{K}\rangle$ and $\langle P_{\eta}\rangle$ in
the Gottfried sum cancel out those terms in $Z=1-\frac{3}{2}\langle
P_{\pi}\rangle-\langle P_{K}\rangle-\frac{1}{6}\langle
P_{\eta}\rangle$, so the value of $\Lambda$ can not be determined
from Eq.~(\ref{gottfried}) or experimental data. It is natural to
assume that the cutoffs are same for $\pi$, $K$ and $\eta$ mesons in
the effective chiral quark model,
$\Lambda_{\pi}=\Lambda_{K}=1500$~MeV~\cite{sw98,Szcz96}, which is
different from the traditional meson cloud model.

In this paper, the parton distributions for mesons are taken from
the parametrization from GRS98 given by Gluck-Reya-Stratmann~\cite{grs98} because the results from
parametrization can be closer to reality than those from models,
\begin{eqnarray}
V_{\pi}(x)=0.942x^{-0.501}(1+0.632\sqrt{x})(1-x)^{0.367},\nonumber\\
V_{u/K^{+}}(x)=V_{d/K^{0}}(x)=0.541(1-x)^{0.17}V_{\pi}(x).
\end{eqnarray}

We also need inputs of constituent-quark distributions $u_{0}$ and
$d_{0}$. But there is no proper parametrization of them because they
cannot be directly measured in the experiment. Therefore, we have to choose
some models as inputs. In this paper the constituent quark model
distributions~\cite{hz81} and the light-cone quark-spectator-diquark
model distributions~\cite{parton} are adopted as two different kinds
of inputs of constituent quark distributions. The constituent quark
model distributions have the following forms
\begin{eqnarray}
u_{0}(x)&=&\frac{2}{\textmd{B}[c_{1}+1,c_{1}+c_{2}+2]}x^{c_{1}}(1-x)^{c_{1}+c_{2}+1},
\nonumber\\
d_{0}(x)&=&\frac{1}{\textmd{B}[c_{2}+1,2c_{1}+2]}x^{c_{2}}(1-x)^{2c_{1}+1},
\end{eqnarray}
where $B[i,j]$ is the Euler beta function, and $c_{1}=0.65$ and
$c_{2}=0.35$ adopted from Ref.~\cite{hz81,kko99} can satisfy the
number sum rules
\begin{eqnarray}
\int^{1}_{0}u_{0}(x)\textmd{d}x=2, ~~~~
\int^{1}_{0}d_{0}(x)\textmd{d}x=1,\label{numsum}
\end{eqnarray}
and the momentum sum rule
\begin{equation}
\int_{0}^{1}xu_{0}(x)\textmd{d}x+\int_{0}^{1}xd_{0}(x)\textmd{d}x=1.\label{momentumsum}
\end{equation}
It is pointed out that there are other different values for $c_{1}$
and $c_{2}$ suggested by Ref.~\cite{hy02}. The light-cone
quark-spectator-diquark model distributions~\cite{parton} are
\begin{eqnarray}
u_{0}(x)&=&\frac{1}{2}a_S(x)+\frac{1}{6}a_V(x),~~\nonumber\\
d_{0}(x)&=&\frac{1}{3}a_V(x).~~
\end{eqnarray}
where $a_D(x) \propto \int [{\rm d}^2 {\bf k}_{\perp}]
|\varphi_{D}(x,{\bf k}_{\perp})|^2$~($D=S$ or $V$) is normalized
such that $\int_0^1 dx a_D(x) = 3$ and denotes the amplitude for the
quark $q$ being scattered while the spectator is in the diquark
state $D$. We adopt the Brodsky-Huang-Lepage prescription~\cite{BHL}
for the light-cone momentum space wave function of the
quark-spectator-diquark
\begin{equation}
\varphi_{D}(x,{\bf k}_{\perp}) =A_{D} \exp
\{-\frac{1}{8\beta^2_{D}}[\frac{m^2_q+{\bf k}^2_{\perp}}{x}
+\frac{m^2_D+{\bf k}^2_{\perp}}{1-x}]\}.
\end{equation}
where ${\bf k}_{\perp}$ is the internal quark transversal momentum,
$m_q$ and $m_D$ are the masses of the quark $q$ and spectator $D$,
and $\beta_D$ is the harmonic oscillator scale parameter. In this
paper we simply adopt $m_q=330$~MeV, $\beta_D=330$~MeV,
$m_S=600$~MeV and $m_V=800$~MeV as often adopted in literature. In
the light-cone quark-spectator-diquark model, the number sum rule
Eq.~(\ref{numsum}) is still satisfied but the momentum sum rule
Eq.~(\ref{momentumsum}) is violated. It should be noticed that in
this paper we adopt the quark-spectator-diquark model, i.e., when
any quark is probed, the other part of the target is served as a
spectator with quantum numbers of a diquark. Thus, some gluon effects
may exist inside the spectators. This means that partial momentum can
be carried by gluons at the initial point within the
quark-spectator-diquark model. Hence there is no need to require the
momentum sum rule Eq.~(\ref{momentumsum}) by quarks as in the
constituent quark model, where the nucleon momentum is distributed
among constituent quarks at the initial point.

\section{ADDITIONAL SYMMETRIC SEA CONTRIBUTIONS}\label{section3}

As is well-known, the quark distributions measured by experiments at
certain $Q^{2}$ include not only non-perturbative intrinsic sea but
also perturbative extrinsic sea~\cite{Bo-Qiang Ma}. Although the
antiquark content in the nucleon mainly comes from the intrinsic
sea, the extrinsic sea should also be considered from a strict sense
when we want to investigate the distributions of quarks and
antiquarks. Therefore, we take into account additional
contribution from the symmetric sea before we use our quark
distributions to compare with experimental data.

We should point out that the quark distribution functions we obtain
in the front can only be proper at a certain $Q _0 ^2$, because the
values of parameters in the model do not evolve according to $Q ^2$.
In the experiment of the E866/NuSea Collaboration, $Q^2$ varies from
about $21$~GeV$^2 / $c$^2$ to more than $160$~GeV$^2 / $c$^2$, so
the change of $Q^2$ should be considered carefully. In this paper,
we choose $Q _0 = 0.7~$GeV/c. For simplicity, we also assume that
the quark distribution functions we get have the same evolution behavior with that of parametrization we adopt,
\begin{equation}
q\left(x,Q^2\right) = q\left(x,Q _0 ^2\right) \frac{q
^{\mathrm{para}}\left(x,Q ^2\right)}{q ^{\mathrm{para}}\left(x,Q _0
^2\right)},
\end{equation}
where $q = u$, $d$, $\bar{u}$, $\bar{d}$. The $q ^{\mathrm{para}}$
stands for the quark distribution of flavor $q$ from
parametrization.

It is found that most parametrizations of parton distributions after
the experiment of E866/NuSea Collaboration had been considerably
affected by it, so we adopt an earlier parametrization, namely CTEQ4
parametrization~\cite{cteq4}. In this paper we will use $\bar{u}$ as
the criterion. Other methods to take the contribution of symmetric
sea may also work, and our method is a feasible choice. As we
can see, the $\bar{u} (x)$ we derived from the model can be larger
than that of CTEQ4 parameterization when $x$ is large, so we should
not add symmetric sea contribution to $\bar{u}$ and $\bar{d}$ any
more when they are larger than the result from parameterization.
Specifically, if $\bar{u} ^{\mathrm{model}} (x)< \bar{u}
^{\mathrm{para}} (x)$, we estimate the symmetric sea
contribution $\delta \bar{u} (x) = \delta \bar{d} (x) = \delta u (x)
= \delta d (x) = \bar{u} ^{\mathrm{para}} (x) - \bar{u}
^{\mathrm{model}} (x)$, otherwise we set the symmetric sea to be
zero. Intuitively, the flavor symmetric sea perturbative extrinsic can be
thought as arising from the splitting of gluons to quark-antiquark
pairs, thus they are flavor symmetric. Similar consideration has also
been adopted to confront the calculated strange and antistrange
distributions with experimental observations in the calculation of
the strange-antistrange asymmetry in the chiral quark
model~\cite{dxm04}.

We compare the parton distributions derived above with the CTEQ4
parametrization and find that $\bar{u}$ and $\bar{d}$ quark
distributions can match well while that $u$ and $d$ quark
distributions really vary. Thus, we will adopt two methods to compare
results with experimental data in the following: (1) we use parton
distributions of both quarks and antiquarks from the model in the
calculation, and this is denoted as ``Model''; (2) we use parton
distributions of $\bar{u}$ and $\bar{d}$ from the model while the
distributions of $u$ and $d$ are parametrization results from CTEQ4
directly, and this is denoted as ``Q.Para.''.

\begin{figure*}
\begin{center}
\resizebox{0.9\textwidth}{!}{%
\includegraphics{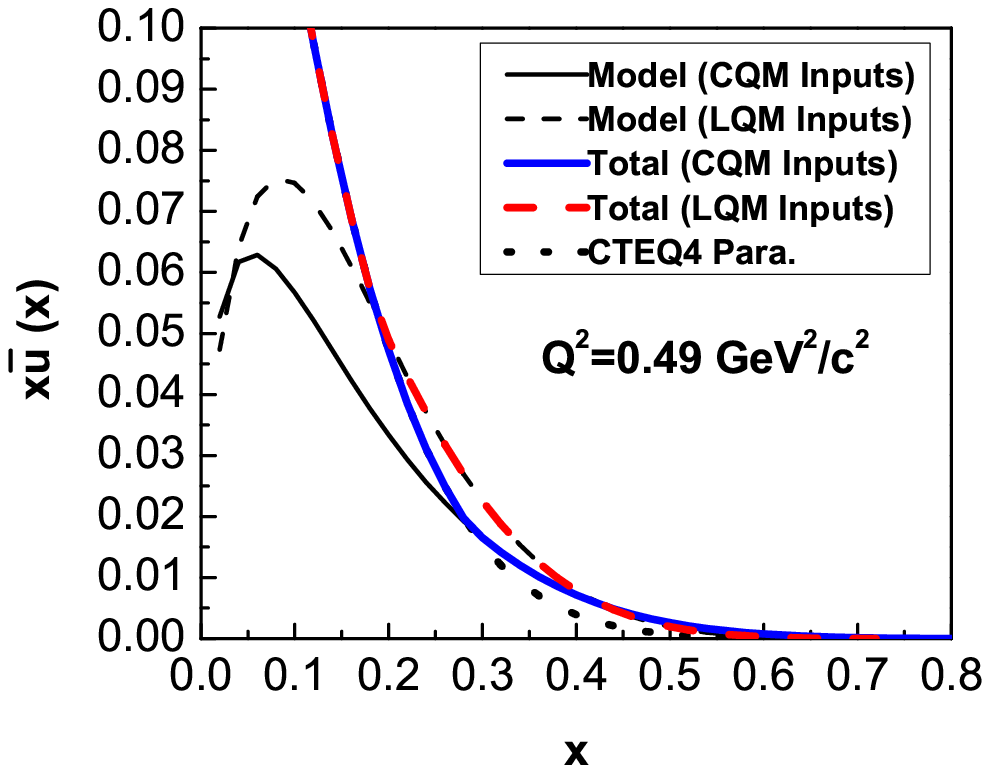}
\includegraphics{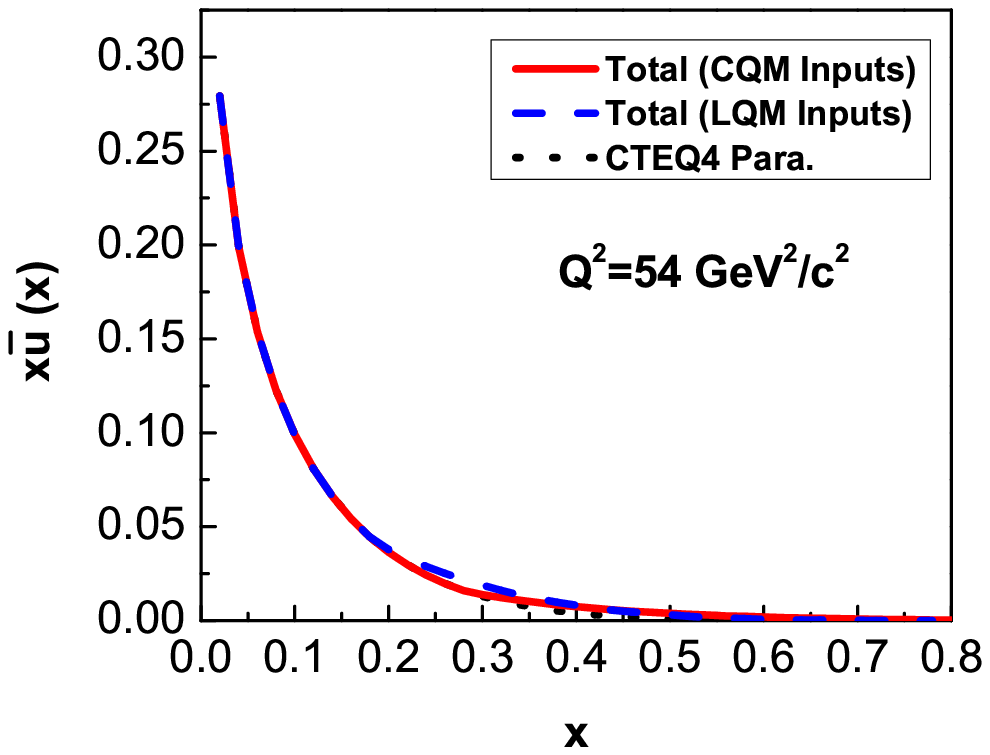}}
\caption{\small Distributions for $x\bar{u}(x)$, with
$Q^{2}=0.49~\mathrm{GeV}^{2}/\mathrm{c}^{2}$ for the left part and
$Q^{2}=54~\mathrm{GeV}^{2}/\mathrm{c}^{2}$ for the right part. The
dotted curve is the result from CTEQ4 parametrization. The thin
solid and dashed curves are the model calculation results with the
constituent quark model (CQM) and the light-cone
quark-spectator-diquark model (LQM) as inputs. The thick
corresponding curves are the corresponding results with chiral quark
model results plus the symmetric sea contributions.} \label{xubar}
\end{center}
\end{figure*}

\begin{figure*}
\begin{center}
\resizebox{0.9\textwidth}{!}{%
\includegraphics{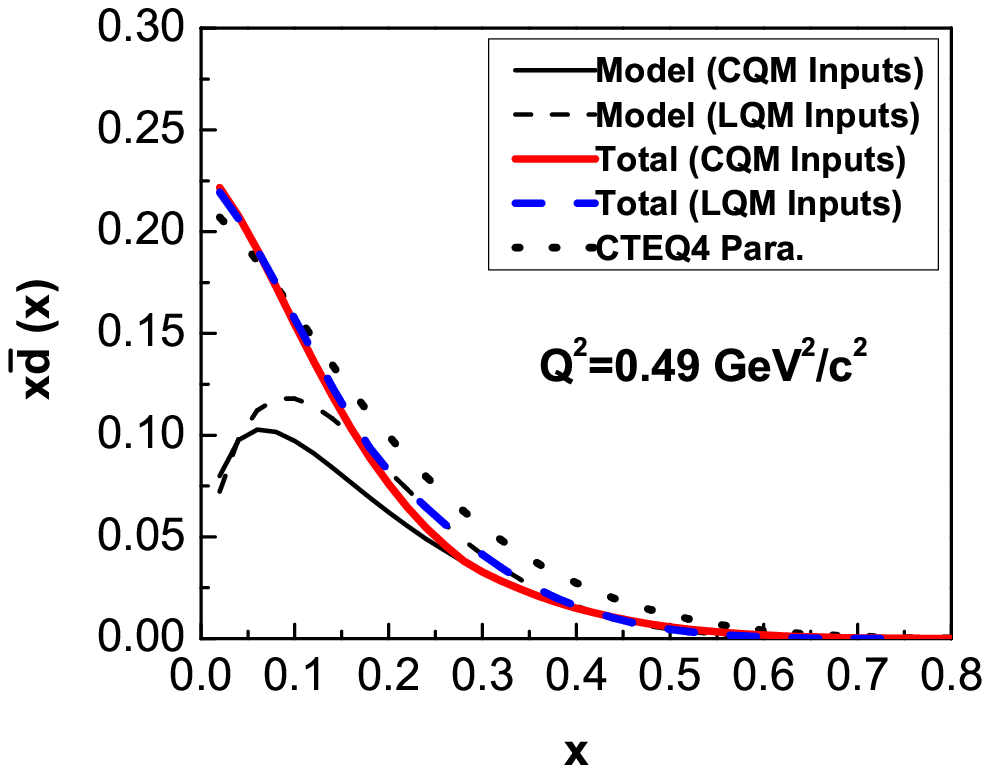}
\includegraphics{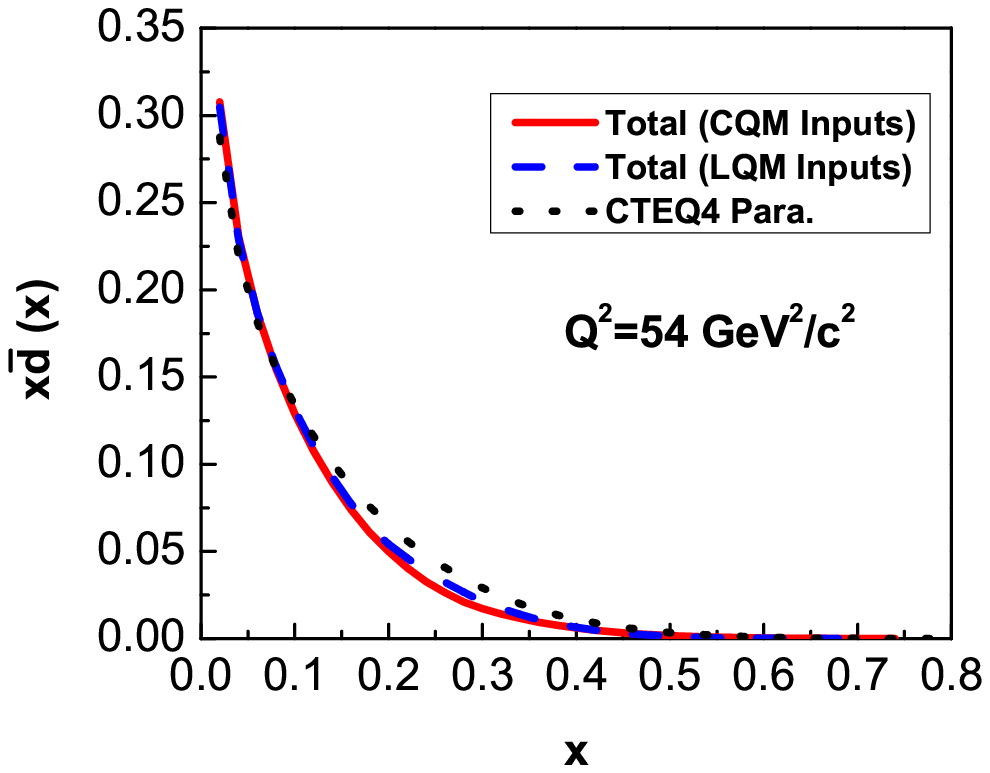}}
\caption{\small Distributions for $x\bar{d}(x)$, with
$Q^{2}=0.49~\mathrm{GeV}^{2}/\mathrm{c}^{2}$ for the left part and
$Q^{2}=54~\mathrm{GeV}^{2}/\mathrm{c}^{2}$ for the right part. The
dotted curve is the result from CTEQ4 parametrization.  The thin
solid and dashed curves are the model calculation results with the
constituent quark model (CQM) and the light-cone
quark-spectator-diquark model (LQM) as inputs. The thick
corresponding curves are the corresponding results with chiral quark
model results plus the symmetric sea contributions.}\label{xdbar}
\end{center}
\end{figure*}
\begin{figure}
\begin{center}
\resizebox{0.54\textwidth}{!}{\includegraphics{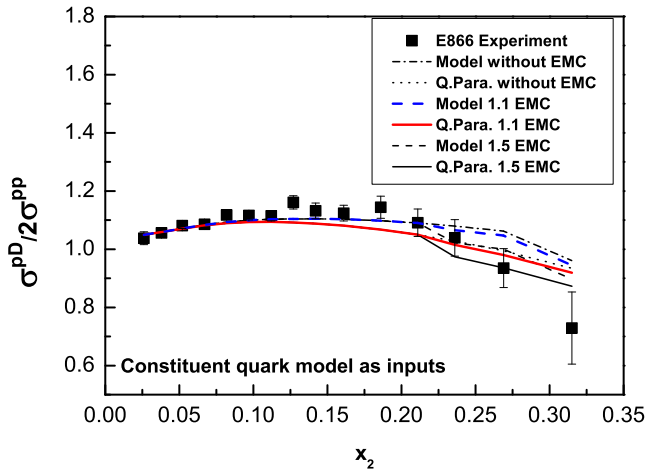}}
\caption{\small The Drell-Yan cross section ratio
$\sigma^{\mathrm{pD}}/2\sigma^{\mathrm{pp}}$ versus $x_{2}$. The
data are taken from E866/NuSea~\cite{E866} experiment. The following
curves are results from model with the constituent quark model (CQM)
as inputs plus symmetric sea contributions. The thin dash-dotted
(dotted) curve is the result that the quark distribution from the
chiral quark model (CTEQ4 parametrization) without EMC effect. The
thick dashed (solid) curve is the corresponding result with EMC
effect for the parameter $\xi = 1.1$. The thin dashed (solid) curve
is the corresponding result with EMC effect for the parameter $\xi =
1.5$.}\label{ratio cq}
\end{center}
\end{figure}

The forms of cross sections we take are
\begin{eqnarray}
\sigma^{\mathrm{pp}} &\propto& \frac{4}{9} u(x_1)\bar u(x_2) + \frac{1}{9}
d(x_1)\bar d(x_2) \nonumber\\&+& \frac{4}{9} \bar{u}(x_1) u(x_2) + \frac{1}{9}
\bar{d}(x_1) d(x_2),
\end{eqnarray}
and
\begin{eqnarray}
\sigma^{\mathrm{pn}} &\propto& \frac{4}{9} u(x_1)\bar d(x_2) + \frac{1}{9}
d(x_1)\bar u(x_2) \nonumber\\&+& \frac{4}{9} \bar{u}(x_1) d(x_2) + \frac{1}{9}
\bar{d}(x_1) u(x_2).
\end{eqnarray}
The influence of heavier quarks is not included as their
contributions can be reasonably neglected. With all these taken into
consideration, we can get the results shown in Fig.~\ref{ratio
diquark} and Fig.~\ref{ratio cq}. We find the results are quite good
except when $x$ is large, regardless of the constituent quark
distribution inputs and the methods we adopt. Accordingly, the
approach taken by us should be reasonable.
\begin{figure}
\begin{center}
\resizebox{0.55\textwidth}{!}{\includegraphics[1,10][329,230]{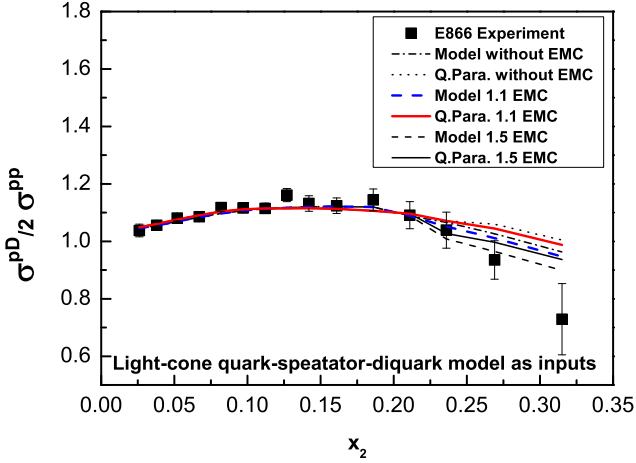}}
\caption{\small The Drell-Yan cross section ratio
$\sigma^{\mathrm{pD}}/2\sigma^{\mathrm{pp}}$ versus $x_{2}$. The
data are taken from E866/NuSea~\cite{E866} experiment. The following
curves are results from model with the light-cone
quark-spectator-diquark model (LQM) as inputs plus symmetric sea
contributions. The thin dash-dotted (dotted) curve is the result
that the quark distribution from the chiral quark model (CTEQ4
parametrization) without EMC effect. The thick dashed (solid) curve
is the corresponding result with EMC effect for the parameter $\xi =
1.1$. The thin
 dashed (solid) curve is the corresponding result with EMC effect for the parameter $\xi = 1.5$.}\label{ratio diquark}
\end{center}
\end{figure}

\section{DISCUSSION ON NUCLEAR EMC EFFECT}\label{section4}

In 1982, the European Muon Collaboration (EMC) at
CERN~\cite{EMCA,EMC} found that the structure-function ratio of
bound nucleon to free nucleon, in the form of
$F^{\mathrm{A}}_2(x,Q^2)/\\F^{\mathrm{D}}_2(x,Q^2)$, is not
consistent with the expectation by assuming that a nuclei is
composed by almost free nucleons with Fermi motion correction taken
into account, and such phenomenon was confirmed by E139
collaboration at SLAC~{\cite{SLAC}}. This discovery, which is called
the nuclear EMC effect, has received extensive attention by the
nuclear and hadronic physics society. There are many models
describing EMC effect
now~\cite{LEST,ME,Jaffe,Carlson,Vary,FEC,RLJ,FECB,ONachtmann}, and a
good review can be found in Ref.~\cite{BLMA}. All these models can
qualitatively describe the data in the mediate $x$ region. The
inclusive deep inelastic scattering (DIS) data are expressed as
$F^{\mathrm{A}}_2(x,Q^2)/F^{\mathrm{D}}_2(x,Q^2)$, which can be
written in the naive parton model as
\begin{equation}
\frac{F^{\mathrm{A}}_2(x,Q^2)}{F^{\mathrm{D}}_2(x,Q^2)}
=\frac{\Sigma_{i}{e_i^2\left[q_i(x,Q^2,\mathrm{A})+\bar{q}_i(x,Q^2,\mathrm{A})\right]}}{\Sigma_{i}{e_i^2\left[q_i(x,Q^2)+\bar{q}_i(x,Q^2)\right]}},
\end{equation}
where $e_i$ denotes the charge of the partons with flavor $i$, and
$q(x,Q^2)$ is the parton distribution function of the nucleon.

\begin{figure}
\begin{center}
\resizebox{0.54\textwidth}{!}{\includegraphics{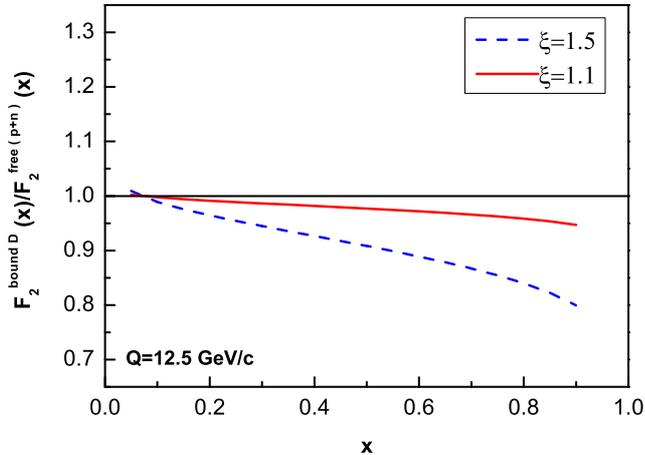}}
\caption{\small The ratio of  bound deuterium
$F_{2}^{\mathrm{D}}(x)$ compared to free
$F_{2}^{\mathrm{p}+\mathrm{n}}(x)$. The solid (dashed) curve is the
result with $\xi = 1.1$ ($\xi = 1.5$). }\label{EMC}
\end{center}
\end{figure}

\begin{figure}
\begin{center}
\resizebox{0.54\textwidth}{!}{\includegraphics{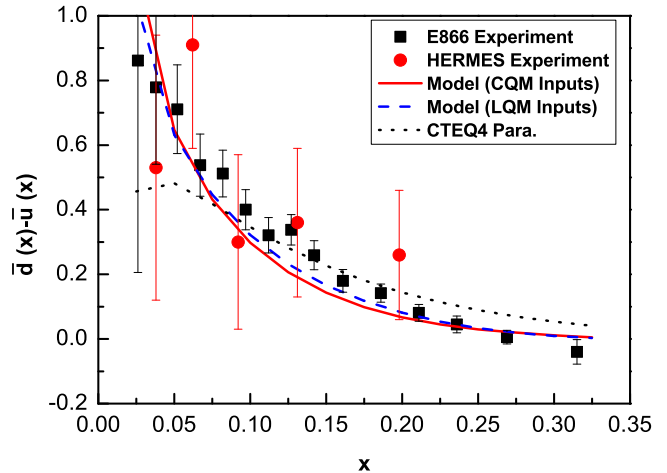}}
\caption{\small The distribution for $\bar{d}(x)-\bar{u}(x)$. The data
are from HERMES~\cite{HERMES} and E866/NuSea~\cite{E866}
experiments. The solid and dashed curves are the model calculation
results in the effective chiral quark model with the constituent
quark model (CQM) and the light-cone quark-spectator-diquark model
(LQM) as bare quark distribution inputs respectively. The dotted
curve is the result from CTEQ4 parametrization. All the values are
scaled to fixed
$Q^{2}=54~\mathrm{GeV}^{2}/\mathrm{c}^{2}$.}\label{minus}
\end{center}
\end{figure}

\begin{figure}
\begin{center}
\resizebox{0.54\textwidth}{!}{\includegraphics{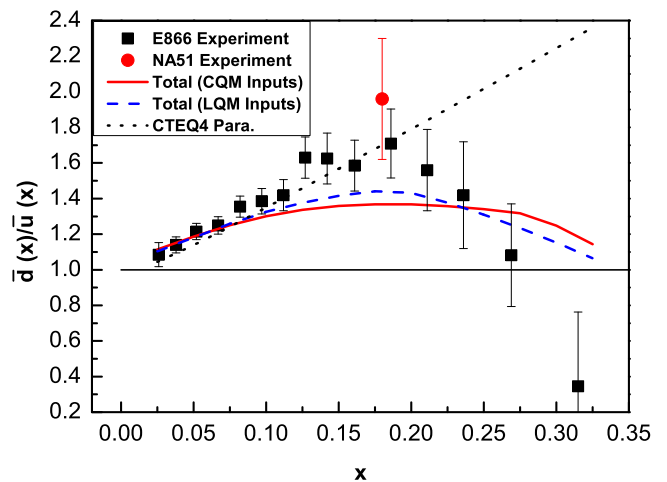}}
\caption{\small The distribution for $\bar{d}(x)/\bar{u}(x)$. The data
are from NA51~\cite{NA51} and E866/NuSea~\cite{E866} experiments.
The solid and dashed curves are the model calculation results in the
effective chiral quark model with the constituent quark model (CQM)
and the light-cone quark-spectator-diquark model (LQM) as bare quark
distribution inputs plus symmetric sea contributions. The dotted
curve is the result from CTEQ4 parametrization. All the values are
scaled to fixed
$Q^{2}=54~\mathrm{GeV}^{2}/\mathrm{c}^{2}$.}\label{divide}
\end{center}
\end{figure}

In the analysis of the E866/NuSea Collaboration, the nuclear effects
in deuterium were assumed to be negligible. As discussed above, the
calculated results of $\sigma^{\mathrm{pD}}/2\sigma^{\mathrm{pp}}$
cannot describe the experimental data well when $x$ is large, and
this is just the region where the nuclear EMC effect may begin to
work. So we will take the EMC effect into account to check how the
results can be changed. In this paper, we choose the so-called
$Q^2$-rescaling model~\cite{Jaffe,FECB,FEC,RLJ}. In this model, the
quark of a bound nucleon in the nuclear medium is considered to have
different confinement size compared with that of the quark in the
free nucleon, and consequently $q ^{\mathrm{A}}(x,Q^2)$ is related
with $q ^{\mathrm{N}}(x,Q^2)$ (the parton distribution in the free
nucleon) by the relation
\begin{equation}
q^{\mathrm{A}} \left(x,Q^2\right) = q^{\mathrm{N}}
\left(x,\xi\left(Q^2\right) Q^2\right).
\end{equation}
At $Q \approx 12.5$~GeV/c, we adopt $\xi = 1.1$ given in the
original work~\cite{FECB}. We show the ratio of
$F_{2}^{\mathrm{D}}(x)$ in deuterium and that in a free proton plus
a free neutron (denoted as p+n) in FIG.\ref{EMC}. We assume
furthermore that the nuclear EMC effect only takes effect when $x
> 0.22$. To show the dependence of the rescaling factor $\xi$, we also adopt a larger value $\xi =
1.5$ as a comparison. Then we can see that the behavior of cross
section ratio at large $x$ is visibly improved. In addition, we find
that the cross section ratio at $\langle x_2\rangle  = 0.315$ is
smaller than 1. We also display $\bar{d}(x)-\bar{u}(x)$ and
$\bar{d}(x)/\bar{u}(x)$ in Fig.\ref{minus} and Fig.\ref{divide}
respectively. As we can see, the behavior of $\bar{d}(x)-\bar{u}(x)$
we derived can match well with the experimental data and
parametrization from CTEQ4, however, $\bar{d}(x)/\bar{u}(x)$ we get
can match with experiment at small $x$ but is not compatible with
the experiment at large $x$, and especially we could not get the
ratio smaller than 1 in the region we consider. It is worthy to
remind that all known models have the result that
$\bar{d}(x)/\bar{u}(x)$ is larger than 1 over all $x$ range.

\begin{figure}
\begin{center}
\resizebox{0.54\textwidth}{!}{\includegraphics{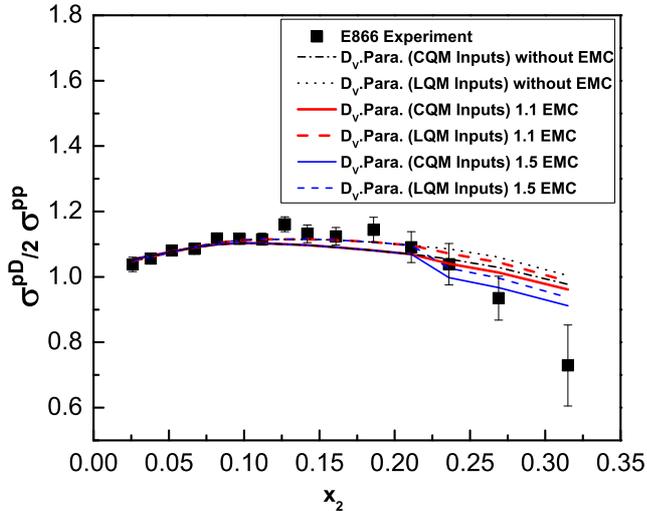}}
\caption{\small The Drell-Yan cross section ratio
$\sigma^{\mathrm{pD}}/2\sigma^{\mathrm{pp}}$ versus $x_{2}$. The
data are taken from E866/NuSea~\cite{E866} experiment. $u(x)$,
$\bar{u}(x)$, $d_{v}(x)$ are obtained by adding CTEQ4
parametrization and symmetric sea contributions. The thin
dash-dotted and dotted curves are the results with $\bar{d}(x)$
derived from the chiral quark model with constituent quark model
(CQM) and light-cone quark-spectator-diquark model (LQM) as bare
quark distribution inputs without EMC effect. The thick solid and
dashed curve are the corresponding results with EMC effect for the
parameter $\xi = 1.1$. The thin solid and dashed curves are the
corresponding results with EMC effect for the parameter $\xi = 1.5$.
}\label{dvpara}
\end{center}
\end{figure}

It is found that we can adopt an alternative procedure to deal with
$u$ quark and $d$ quark to compare the result with the experiment.
We assume $u(x) = u^{\mathrm{para}}(x)$ and $d(x) = d_v(x)
+\bar{d}(x)$, where
$d_v=d^{\mathrm{para}}-{\bar{d}}^{\mathrm{para}}$ is the valance $d$
quark distribution, and $\bar{d}$ is the result we get from the
chiral quark model with symmetric sea quark distribution added. The
result is denoted as ``$\mathrm{D}_{\mathrm{v}}.\mathrm{Para}.$''
and displayed in Fig.~\ref{dvpara}. We find that this result can
also match well with the experimental data.

So we suggest a procedure for experimental data treatment to derive
$\bar{d} (x)/\bar{u}(x)$ and $\bar{d}(x)-\bar{u}(x)$ from the cross
section ratio $\sigma^{\mathrm{pD}}/2\sigma^{\mathrm{pp}}$: first
assume $u(x) = u^{\mathrm{para}}(x)$, $\bar{u}(x) = {\bar{u}}^
{\mathrm{para}}(x)$ and $d(x) =
d^{\mathrm{para}}(x)-{\bar{d}}^{\mathrm{para}}(x) +\bar{d}(x)$, then
the experimental data is used to fit the behavior of $\bar{d}(x)$.
This procedure might give results more compatible with model
predictions than the method adopted in the E866/NuSea analysis.


\section{RESULT AND CONCLUSION}\label{section5}

In this work, we calculate the light flavor quark and antiquark
distributions within the effective chiral quark model by using the
constituent quark model and the light-cone quark-spectator-diquark
model as inputs respectively, and revise the results by taking
into consideration the symmetric nucleon sea contributions, the $Q
^2$-evolution and the nuclear EMC effect. The distributions of
$\sigma^{\mathrm{pD}}/2\sigma^{\mathrm{pp}}$ and $\bar{d}(x) -
\bar{u}(x)$ match with the experimental data, and $\bar{d}(x) /
\bar{u}(x)$ is compatible with the experiment at small $x$, while
the behavior of $\bar{d}(x) / \bar{u}(x)$  in large $x$ region is
different from the experimental result. However, the result directly
measured in E866 experiment was only
$\sigma^{\mathrm{pD}}/2\sigma^{\mathrm{pp}}$, whereas
$\overline{d}(x)/\overline{u}(x)$ was derived indirectly from
$\sigma^{\mathrm{pD}}/2\sigma^{\mathrm{pp}}$ with several
assumptions. Although it is entirely possible that the analysis of
E866/NuSea Collaboration was correct, it is also possible that the
analysis of them was based on some excessive assumptions. For
example, the assumption of $\bar{u}(x)+\bar{d}(x)$ fixed as
parametrization may have a strong influence on the ratio
$\bar{d}(x)/\bar{u}(x)$ derived from data of cross sections. In
addition, it is worth noting that the nuclear EMC effect should be
considered carefully when $x$ is large. Therefore, without ruling
out these possibilities and considering all we discussed before
carefully, the results of $\bar{d}(x) / \bar{u}(x)$ extracted from
other quantities might be not so reliable as the results of
$\sigma^{\mathrm{pD}}/2\sigma^{\mathrm{pp}}$. We also suggest an
alternative procedure to derive $\bar{d}(x) / \bar{u}(x)$ and
$\bar{d}(x) - \bar{u}(x)$ from the experimental data of
$\sigma^{\mathrm{pD}}/2\sigma^{\mathrm{pp}}$. Thus it is important
that more precision experiments should be carried out to enable more
direct and accurate determination of sea quark and antiquark
distributions.

\section*{Acknowledgement}
This work is supported by National Natural Science Foundation of
China (Nos.~11021092, 10975003, and 11035003), and National Fund for
Fostering Talents of Basic Science (Nos.~ J1030310, J0730316). It is
also supported by Principal Fund for Undergraduate Research at
Peking University.

\end{document}